# Network Topology and Subgap Resonances Observed by Fourier Transform Scanning Tunneling Microscopy in Cuprate High-Temperature Superconductors

J. C. Phillips

Dept. of Physics and Astronomy,

Rutgers University, Piscataway, N. J., 08854-8019

## ABSTRACT

Fourier transform scanning tunneling microscopy on BSCCO subgap resonances has deciphered an octet of "quasi-particle" states that are consistent with the Fermi surface and energy gap observed by ARPES, but the origin of the high-intensity **k**-space octets and the sharply defined **r**-space checkerboard is unexplained. The filamentary ferroelastic nanodomain model that *predicted* the **r**-space checkerboard also explains the **k**-space octets and the origin of the apparent anisotropic surface d-wave gap by using strong electron-phonon interactions outside the $CuO_2$ planes. The topological model identifies the factors that stabilize high-intensity **k**-space octets in the presence of a very high level of irregular **r**-space checkerboard noise.

## 1. Introduction

Recent high-resolution **r**-space tunneling (Pan et al 2001; Lang *et al.* 2002) and **k**-space photoemission (Bogdanov *et al* 2000; Johnson *et al.* 2001; Lanzara *et al.* 2001; Gromko *et al.* 2002) experiments on the spatial and momentum electronic structure of high-temperature superconductors (HTSC) have revealed very irregular and anisotropic features with spectral and temperature dependencies closely related to the superconductive energy gap. So far most theoretical efforts to interpret these observations (Shen and Schrieffer 1997; Laughlin 1999; Allen 2001) have relied on the effective medium approximation (EMA) (Fermi surfaces, Fermi liquid scattering theory, Mott-Hubbard models, etc.), presumably on the grounds that the techniques most often used to describe superconductive metals (such as Al, Pb, Nb, $V_3Si$, etc.) are based on this approximation. Suspiciously, *none* of the many widely studied, very *weakly* disordered materials, for which these methods are known to be at least partially successful, are HTSC. Moreover, ceramic cuprate HTSC have pseudoperovskite crystal structures, and it has long been known that the perovskite family of materials is characterized by exceptionally large spatial inhomogeneities (*strong* disorder) on a wide range of length scales (Phillips and Jung 2001). Thanks to the tunneling studies, we know that the smallest planar scale of these spatial inhomogeneities is about 3 nm or even less, which is comparable to the planar superconductive coherence length. (Even before these experiments, the smallest scale for other perovskites, such as $BaTiO_3$, had already been reduced to 5 nm.) When one adds to this the fact that the metallic character is made possible *only by doping*, and that the dopants themselves can segregate or self-organize to form percolative paths on similarly small (or even smaller) scales, the suitability of perturbative EMA and Fermi surface constructions for theoretical modeling of high-resolution data becomes extremely dubious (Mueller *et al.* 1987; Gorkov and Sokol 1987, Phillips 1987, 1990, 1995; Phillips and Jung 2001; Mueller 2002). Such self-organization improves dielectric screening of internal electric fields, as observed explicitly for clustered ferromagnetic semiconductor impurities (Timm *et al.* 2002).

The general principle that guides the construction of theoretical models of complex systems is economy of means, that is, one seeks not merely the simplest model, but rather the simplest model that is compatible with experiment, and that principle has determined the model discussed in this paper. However, the experimental data base now appears to *require* a nanostructural model that goes far beyond the simplistic homogeneous single-phase postulated by effective medium models. Another way of saying this (Phillips, 1987, 1990; Gorkov and Sokol 1987), even without the most recent data, is that HTSC is not a simple phenomenon, and it is not reasonable to expect to be able to explain it with minor phenomenological modifications of effective medium theories.

There certainly are many departures from simple s-wave BCS superconductivity evident in a wide range of experiments that are indicative of anisotropic effects, and if one is committed to interpreting data entirely within the EMA, d-wave anisotropy of the energy gap at the Fermi line is not a bad place to begin. The problem is that the inadequacies of a partial wave expansion of positive energy states in vacuum become apparent very quickly. For example, there is often a tail in the density of states observed at negative energies below the gap; this tail is indicative of the existence of subgap states, and in optical experiments, such as ARPES, there seem to be many more of these in the (110) directions than in the (100) in the superconducting state (Phillips 2003) (preceding paper). It is natural to attempt to describe this angular anisotropy with the smallest possible value of $l$, $l = 2$ (d waves). However, when this is done one immediately encounters a serious problem. The density of states that is observed experimentally should *always* exhibit a *larger* tail, due to inhomogeneous broadening, than is predicted (Won and Maki 1994), but in practice the tail is often observed to be *smaller*; in rare cases the allowable gap anisotropy is even reduced from 100% to 3% (Shimada *et al*. 1995). This failure of a *rigorous* mathematical limit should logically have led to the abandonment of either the EMA or the d wave concept. Instead, it led mainly to the proliferation of EMA "fixes", such as d + s waves, etc., a merely descriptive approach that eventually led nowhere.

The experimental solution to this problem is to find a way of studying the subgap states in more detail. Within the EMA one would do this by improving the resolution, both energetic and angular, of ARPES. The results of such improvements are discussed in (Phillips 2003). Strong evidence has been found (Bogdanov *et al.* 2000) for extremely rapid angular variations of peak widths $2\Gamma(E)$, sometimes on a scale of $< 1°$. To explain such strong angular variations within the EMA requires a value of $l \sim 100 >> 2$, which is quite unreasonable. Thus one must go outside the EMA, and study variations in $(E, \mathbf{r})$ as well as variations in $(E, \mathbf{k})$, because $\mathbf{k}$ is a good quantum number only in the EMA. With STM one can observe spatial variations of this kind, but at first sight the variations appear to be noisy and completely unrelated to the Fermi line, as they occur on a length scale ten times that of the unit cell. However, ingenious Fourier analysis of dynamical pair distribution functions $P(E, \mathbf{r}_1, \mathbf{r}_2)$ has revealed (McElroy *et al.* 2002) a rich internal structure in the Fourier transform of $Q(E, \mathbf{r}_1 + \mathbf{r}_2, \mathbf{r}_1 - \mathbf{r}_2)$ for $E < E_g = 40$ meV. [Note that $\mathbf{k}$ is conjugate to $(\mathbf{r}_1 + \mathbf{r}_2)$, and $\mathbf{q}$ to $(\mathbf{r}_1 - \mathbf{r}_2)$]. This structure is concentrated near $k_F$, and it shows an unexpectedly sharp angular dependence (on a scale of 5°, corresponding to $l \sim 10$), similar in that sense to the ARPES data. Similar sharp angular dependencies are known in the dispersive LO phonon anomalies, measured by neutron scattering, reviewed (Phillips 2001, Phillips and Jung 2002b) in the context of the filamentary model.  Attempts to explain the ARPES and STM data solely in the context of electronic EMA Fermi line d-wave models cannot account for the anisotropically gapped phonon structure, for which data are becoming increasingly convincing (Pintschovius *et al.* 2002).

  The most recent high-resolution STM experiments (McElroy *et al.* 2002) are the main topic of this paper.  The studies encompass an awesome data base:  a 65 nm field of view encompassing 40,000 unit cells is scanned with 0.13 nm resolution at energy intervals of 2 meV between the limits ± 30 meV.  Fourier transform STM (Petersen *et al.* 1998) leads to the determination of the momentum distribution of a fraction of subgap states (that is, the observed structure is superimposed on a noisy background).  These subgap states behave as wave packets that are localized near eight correlated small parts of the Fermi surface, indexed by $\mathbf{k}_i$, and which scatter strongly only among themselves with scattering vectors $\mathbf{q}_{ij} = \mathbf{k}_i - \mathbf{k}_j$.  The $\mathbf{k}$ components of the octet wave packets satisfy a gap equation,

$$E^2(\mathbf{k}) = \varepsilon^2(\mathbf{k}) + \Delta^2(\mathbf{k}) \tag{1}$$

commonly obtained for any collective two-band model.  The gap $\Delta(\mathbf{k})$ shows d-wave anisotropy, is zero in (11) directions, and has a maximum value near $\Delta_0 = 40$ meV in the (10) direction.  Because it resembles the d-wave gap reported in selected early ARPES experiments, and because of the electron-hole symmetry implied by (1) (and confirmed experimentally for the wave-packet states), the gap is identified as superconductive. (McElroy *et al.* 2002) suggest that the octet structure in $\mathbf{q}_{ij}$ is caused by peaks in the joint density of initial and final states $n_i(\omega)n_f(\omega)$ associated with crescent- (or banana-) shaped sections of the bulk Fermi surface.  Notice that in regions of well cleaved samples there is a polygonal (here idealized as checkerboard) large/small gap pattern (Pan et al 2001; Lang *et al.* 2002), and it is such cases that we discuss here.  Examples where this pattern is not observed (Howald *et al.* 2002) are much more complex, and interpretation of the electronic structure in such cases lies beyond the framework of the present discussion.

At first it appears that the density of states mechanism is sufficient to explain the octet structure, as with $n_i(\omega)n_f(\omega) \sim 1/(\omega - \Delta) \sim (q\xi)^{-2}$ for $q\xi > 1$ [here $\xi$ is the coherence length $\sim 3$ nm, and $q \sim \omega$], one obtains Lorentzian peaks. However, the most recent data (McElroy *et al.* 2002) exhibit more sharply defined Gaussian peaks. More importantly, the integrated intensity of the observed octet peaks appears to be about 10% of the total signal. The path length of a percolating wave packet in the 65 nm field of view should be about 75 nm, even assuming shortest path (not random walk) classical percolation in two dimensions (Phillips 2001b). With a coherence length of 3 nm, this implies a relative signal intensity of $e^{-25}$. Thus the observed intensity is more than $10^6$ larger than expected from ordinary scattering theory. Note also that while such anomalous coherence is enhanced in the superconductive state by the density of states singularity at $\omega = \Delta$, similarly large anomalous coherence has been observed both in microwave (Kusko *et al.* 2002) and ellipsometric infrared (Bernhard *et al.* 2002) experiments even in the normal state.

Normally one assumes that scattering resonances in energy are caused by either density-of-states factors, or by matrix elements, while scattering resonances in momentum are more likely to be geometrical in nature. Thus the coincidence of the coherence length with the nanodomain diameter suggests a geometrical resonance associated with percolation of nearly stationary states associated with the checkerboard pattern on the length scale of 3 nm. The filamentary model envisions the motion of the octet wave packet envelope in real space, and it enables us to construct the needed microscopic mechanisms that describe such resonances. Many discussions have considered only EMA models for HTSC, because these seem to be so much simpler than percolative models (Phillips, 1987, 1990; Gorkov and Sokol 1987; Mihailovic *et al.* 2002). At the same time the EMA models may not explain the origin of the octet wave packet states, the self-scattering property, or even the checkerboard large/small gap pattern, also *predicted* long ago (Phillips 1990; Goodenough and Manthiram 1990). Thus now is a good time and here is a good place for introducing novel percolative geometrical ideas to describe correlations on length scales much larger than unit cell dimensions.

## 2. Simple Examples: Chains and Planes, Checkerboards

The filamentary concept enables us to address one of the oldest problems in cuprate phenomenology: is it the chains or is it the planes that cause HTSC in YBCO? Dramatic and beautiful results have been obtained (Derro *et al.* 2002) for the local energy gap on the surface chains of YBCO by scanning tunneling microscopy. The authors *assume* that the Cooper pairing interaction is confined to *stiff* $CuO_2$ planes, and in the spirit of the EMA ascribe the 25 meV superconducting gap $\Delta$ observed in the *soft* surface chains to a proximity effect involving coherent c-axis dispersion that induces smaller gaps in soft chains from larger gaps in stiff planes.

In continuum theories proximity effects are induced (Morr and Balatsky 2001) by interlayer kinetic energies described by a one-electron bandwidth parameter $t_\perp$. One then obtains the chain gap $\Delta_c$ by solving the gap equation numerically, but the obtained solution is fitted nicely and simply by $\Delta_c = (t_\perp/t_{||})\Delta^*$, where $\Delta^*$ in the $CuO_2$ planes is estimated (Morr and. Balatsky 2001) to be 40 meV. This estimate of $\Delta^*$ is consistent with the maximum (not average) values of $\Delta$ obtained in ARPES data and with STM data for BSCCO (Pan et al 2001). The c axis band width $t_\perp = 0.12$ eV is 40% of the planar band width $t_{||} = 0.3$ eV, and is taken from old ARPES data (Schabel *et al.* 1998) ; a much better estimate of bilayer splitting is obtained from high resolution momentum spectra (Chuang *et al*.2001) on *overdoped* BSCCO: $t_\perp = 0.055$ eV. With $\Delta^* = 40$ meV, and $t_\perp/t_{||} = 0.2$, then an honest estimate is $\Delta_c = 8$ meV, which is 3x too small. Moreover, the energy scale of the superconducting gap observed ellipsometrically in the infrared is *much larger* for chain axis polarization (Bernhard *et al.* 2002), which, as they say, "is hardly consistent with a proximity induced SC state in the chains". In other words, by trying to force the tunneling data to fit a perturbative plane-source continuum model, (Derro *et al.* 2002) have reached qualitatively incorrect conclusions that are inconsistent with independent anisotropic infrared data that should be qualitatively correct, as the anisotropy is measured directly.

In the PJ filamentary model the chains are connected coherently to the planes through strongly disordered resonant tunneling centers. The Cooper pairing interaction was *predicted* to be largest either at these tunneling centers, or in the chains, and to be *small in the planes*. The $\Delta_c = 25$ meV gap is not *induced* in the chains by the proximity effect, it is *native* to them, in agreement with (Bernhard *et al.* 2002). The chain gap is caused by poorly screened giant electron-phonon interactions near chain segment ends, where there are strongly disordered resonant tunneling centers in the BaO layer that connect the chains to the planes. In this *predictive* model the large local value of $\Delta_c \sim 25$ meV occurs in the chains, and the local value of $\Delta_p \sim 0 - 8$ meV in the planes is small, as expected from electron-phonon interactions, because the planes are stiffer than the chains, and a weak e-p interaction can be cancelled by strong Coulomb interactions. (In other words, it would be better to say that the chains induce superconductivity in the planes, than vice-versa, and replace the relation $\Delta_c = (t_\perp/t_\parallel)\Delta^*$ with $\Delta^* = (t_\perp/t_\parallel)\Delta_c$, although neither continuum statement is correct topologically.) The *predictions* of filamentary theory are quantitatively consistent with both the tunneling and ellipsometric infrared data, while unreasonably large values of $\Delta^*$ and $t_\perp/t_\parallel$ are required to sustain the plane source model.

The question of *where* the electron-phonon interaction is strong, and where it is weak, is crucial to constructing a microscopic theory of HTSC. In the filamentary model the strong electron-phonon interactions occur outside the $CuO_2$ planes, and path segments in the latter function simply as weak links between the strongly superconductive path segments that lie *outside the planes*. As we will see in Sec. 4, it is the weakness of the electron-phonon interaction in the $CuO_2$ planar links that makes it possible to observe the octet "quasi-particle" structure, and it is also this weakness that causes the energy gap to appear to be anisotropic with d-wave functional behavior.

The *predictions* of the filamentary nanodomain model (Phillips and Jung 2001) concerning the weakness of superconductivity in the $CuO_2$ planes and the existence of a checkerboard pattern of 60 meV pseudogaps alternating with weakly superconductive nanodomains explains, *without further assumptions*, the very recent and unexpected STM data on $CuO_2$ planar spectra of metastable $CuO_2$ terraces on BSCCO (Misra *et al.* 2002). They indeed observe a 60 meV gap, with a rather broad peak in the density of states, very similar to the "blue" pseudogaps observed in STM on BSCCO BiO natural cleavage planes (Pan et al 2001; Lang *et al.* 2002). They see *no* evidence in the $CuO_2$ planes for a narrow "red" superconductive gap peak near 40 meV. Instead, there appears to be a 10 meV insulating region around $E_F$, which the authors attempt to explain with a *tunneling* model in which **k** is assumed, for no reason, to be a good quantum number. All this structure is fully consistent with the filamentary nanodomain model if one simply assumes that the resonating tunneling center states that connect the merely marginally stable outer $CuO_2$ plane to the next BiO plane are Jahn-Teller split with a low-energy cutoff in the distribution of splitting energies around 10 meV. The latter is a reasonable cutoff, as there are many acoustic zone-boundary phonons at this energy. (Incidentally, this energy can also be used as an upper cutoff to estimate relaxation energies involved in isotope shifts, for example. It is large enough to explain, at least in general terms, the smallness of isotope shifts near optimal doping.) The central point of their data – the completely unexpected absence of superconductivity in an isolated $CuO_2$ plane – is one of the *basic assumptions* of the filamentary nanodomain model (Phillips and Jung 2001) dating back more than 10 years (Phillips 1990); by any reasonable standard, the observed non-metallic and non-superconductive properties of the surface $CuO_2$ plane are a spectacular *predictive* success of the topological theory.

The completely unexpected absence of superconductivity in an isolated $CuO_2$ plane observed by tunneling into metastable $CuO_2$ terraces appears to be in conflict with a recent report (Balestrino *et al.* 2002) of superconductivity at 60K in a single $(CaCuO_2)$ "block" sandwiched between two $(Ba_{0.9}Nd_{0.1})CuO_{2+x}$ charge reservoir blocks grown by pulsed laser deposition. Even if one ignores the very relevant question of how close to an ideal monolayer one can expect to grow a film by pulsed laser deposition, it is still important to note that the ideal single "block" consists of $(CaCuO_2)_2$ *bilayers*, in other words, *two* $CuO_2$ planes. Depending on the (unknown) effects of pulsed laser deposition and misfit stresses with the charge reservoir blocks, the intervening Ca plane could easily contain a few % interstitial ("apical") O ions which would function as centers of strong interplanar electron-phonon coupling, as supposed in the filamentary model.

The alternating (checkerboard) large/small gap pattern (Pan et al 2001; Lang *et al.* 2002) is, of course, irregular, with a correlation length of $\xi \sim 1.5$ nm, corresponding to a nanodomain diameter of 3 nm. This diameter appears to be constant (*independent of* *doping*), and it must be an *intrinsic* property of the material. (It cannot be caused by surface impurities, or by bulk impurities such as Ni, as these cannot explain the constant diameter, would not create pseudogaps and cannot explain the abrupt gap crossover between 40meV in the "red"squares, and 60 meV in the "blue" squares. Moreover, Ni impurities are magnetic, and if the checkerboard pattern were somehow to be stabilized by Ni impurities, they would be centered in the AF "blue" squares. In fact, they are concentrated mainly in the "red" squares (Lang *et al.* 2002). This is just what one would expect if the pseudogap is formed by Jahn-Teller distortions of $CuO_n$ polyhedra, with the latter distortions being suppressed by Ni impurities. Or one could simply observe that the solid solubilities of impurities in metals are always much larger than in semiconductors.) The smaller and much narrower superconductive gap $\Delta \sim 40$ meV in the "red" squares (Lang *et al.* 2002) is the one that determines energy scales for the octet, while the larger and much broader gap in the "blue" squares appears to be a pseudogap with $\Delta_{ps} \sim 60$ meV.

It is important to have some idea of the physical mechanism responsible for the "red-blue" checkerboard pattern, as it dominates the $\mathbf{r}$ – space patterns. Several explanations have been proposed (Lang *et al.* 2002), but only one of them contains a physical mechanism and actually *predicted* the checkerboard pattern, and that is the ferroelastic nanodomain model (Phillips 1990; Goodenough and Manthiram 1990; Phillips and Jung 2001). In this picture the pattern is driven by Jahn-Teller distortions, it relieves stresses created by internal elastic misfit, and it is *unrelated to the $\mathbf{k}$-space Fermi surface*. There are two separate and distinct phases, and by balancing their prototypical lattice constants, most of the misfit stress is relieved (Fogel *et al.* 2002). Incidentally, the 4x4 checkerboard structure observed around vortices (Hoffman *et al.* 2002) is smaller because the vortex magnetic field alters the energy balance between the two phases through magnetoelastic coupling, which is strong in perovskites (Fiebig *et al.* 2002). This interaction is irrelevant to the competition between Jahn-Teller and Cooper pairing interactions that determines percolative effects in the intermediate phase at $\mathbf{H} = 0$.

The effect of the checkerboard pattern on subgap wave packet dynamics in the $CuO_2$ plane is easy to visualize. The "red" wave packets are specularly reflected by [10] or [01] (Cu-O-Cu) "red-blue" interfaces, as in classical Sinai billiard models of percolation (Polyakov *et al.* 2001). The multiple reflections within a nanodomain create a nearly standing wave pattern that resembles a Wannier function (Marzari and Vanderbilt 1997) subject, of course, to the exclusion principle for partially filled bands. This "red" localized function coherently overlaps a similar wave function in a nearly diagonally adjacent "red" nanodomain, which is a basic process that can be referred to as "quantum percolation" (Phillips 1990).

The 65 nm field of view (McElroy *et al.* 2002) corresponds to a very large 22 x 22 checkerboard of 3 nm nanodomains, so this basic process is repeated many times as a given state percolates coherently across the area. As a result of many multiple reflections, the nanodomain interfaces, although locally irregular, because of the central limit theorem effectively reach a nearly mean-field geometry that resembles an almost regular checkerboard, as in Fig. 1(a). The average interface is oriented parallel either to [10] or to [01]. The first correction to this topology is to allow the blue squares to alternate in size, as in Fig. 1(b). This alternation deforms the red squares into rectangles of both x and y longer axis orientations. In this way one generates red nanodomains of orthorhombic symmetry, rather than tetragonal symmetry. Note that the red nanodomains alternate in (x,y) orthorhombic orientations when the overall symmetry is tetragonal, and will continue to do so even if the overall symmetry is orthorhombic, providing that the local orthorhombicity is larger than the macroscopic orthorhombicity. (Haskel *et al.* 1996) have shown with EXAFS that the range of the phase diagram containing the intermediate HTSC phase coincides rather well with the range of the phase diagram where the local orthorhombicity is larger than the macroscopic orthorhombicity and the latter is non-zero (corresponding to what is called directed percolation) (Phillips 1999b). Technically speaking, these geometrical aspects of nanodomain patterns reflect nonlinear textures created by ferroelastic constraints (Rasmussen *et al.* 2001) that suppress the accumulation of orthorhombic misfit energies.

### 3. Octet Resonances

The question now arises as to how the **k** -space octet states are formed in the presence of the **r** – space checkerboard. The actual values of $k_i$, $k_f$, and **q** are contained in superpositions of the amplitudes of many filaments and appear to be hidden in the statistics of the Fourier STM deciphering.  At optimal doping the sizes of the two kinds of squares are approximately equal, and this implies that it should be possible for the superconductive states to percolate along "red" squares from corner to corner.   The observed $\mathbf{q}_{ij}$ depend on $\mathbf{k}_i$ and imply that octets determined by $\mathbf{k}_i$ represent very good approximations to the red nanodomain Wannier functions of  E (or $\mathbf{k}_i$) selected filaments. These octets consist of two orthorhombically dual quartets, ($\pm k_x$, $\pm k_y$) and     its orthorhombic counterpart ($\pm k_y$,   $\pm k_x$).   (McElroy $et\ al.$ 2002) suggest that the octet structure in $\mathbf{q}_{ij}$ is caused by peaks in the joint density of initial and final states $n_i(\omega)n_f(\omega)$, but this mechanism does not explain the observed dual orthorhombic symmetry or even the formation of the peaks in the presence of strong disorder.  In the present geometrical resonance model the first quartet is generated by multiple reflections within percolatively connected orthorhombic red squares of one (x,y) orientation, while the second one refers to red squares with the opposite (y,x) orientation.  [Alternatively the E-based selection could project separate filaments of both orthorhombic symmetries.]  The values of $\mathbf{k}_i$ are preserved but are (x.y) reversed between the two orientations because this is the (broken) symmetric condition for maximal coherent overlap in the corner joints between adjacent nanodomains with reversed (x,y) orientations, as well as maximum dual conductivities. It is striking that of all the seven allowed scattering vectors, narrow scattering resonances are observed for six $\mathbf{q}_i$ (i = 1,…,7) but not for $\mathbf{q}_4$ = -2$\mathbf{k}_F$.  This suggests destructive scattering between superconductive percolating Cooper pairs and either charge or spin density wave percolation with $\mathbf{q}$ = $\mathbf{q}_4$ at the interfacial boundaries between the "red and blue" squares.

The physical picture so far is one of superconductive wave packets with two correlation lengths.  The first length $\xi \sim 1.5$ nm refers to nanodomains; it includes the exponential background in the q plots (McElroy $et\ al.$ 2002), and it produces average scattering angles of 5° (Pan et al 2001).   These small scattering angles are in good agreement with the observed narrow $\mathbf{q}$ peak widths, and they are not directly dependent on the joint density of states.  The second length is very long (> 30 nm) and it generates the octet resonances by the multiple specular reflection mechanism (coherent projection (Manoharan *et al.* 2000)) described above. The percolation paths are determined by the dopants, and each filament is indexed by its own ($\pm k_x$, $\pm k_y$). Nevertheless, the octet values of $\mathbf{q}_{ij} = \mathbf{q}$ are derived from the same formula for all filaments, because they all percolate through the checkerboard maze, and $\mathbf{q}$ is determined by the *average* geometry of the maze (central limit theorem). If a filamentary segment is too short, that is, if it does not percolate across a large fraction of the field of view, then it does not contribute to the $\mathbf{q}$ signal, and it is part of the background noise. There is no evidence for low-energy transport in the pseudogap "blue" squares, but there is indirect evidence that the pseudogap is associated with charge or spin density waves that are antiparallel mixtures of $\mathbf{k}_i$ and $\mathbf{k}_i + \mathbf{q}_4 = \mathbf{k}_f$. If the path passes through a "red-red" interface too close to the nodal direction $\theta = \pi/4$ it becomes, in effect, a normal-state filament. The normal-state filaments exhibit anomalous behavior of their quasiparticle peak widths $2\Gamma(E)$ (Bogdanov *et al.* 2000, Phillips 2003) that is suggestive of very weak electron-electron scattering.

One might worry that the octet of one filament could scatter destructively off the octet of another filament (random phase approximation). This may happen in a very narrow cone near the nodal direction $\theta = \pi/4$, where the orthorhombic counterparts interfere, but in general such destructive interference appears to be weak. The natural reason for this is the proximity of the intermediate phase to the metal insulator transition, which renders the filamentary network dilute. Moreover, the normal-state conductivity is maximized by dopant configurations that minimize interfilamentary scattering. Indeed, when the dopant density n exceeds the percolation density $n_c$ by about a factor of two, interfilamentary scattering does cause a first order transition to the non-HTSC Fermi liquid state. This is a general property of filamentary intermediate phases (Phillips 2002b). Differential sensitivity to O and Sr dopants observed in experiments with LSCO epitaxial films (Bozovic *et al.* 2002) clearly lies outside the range explicable by the EMA, but it is natural for self-organized filaments, as the O mobility is much larger than the Sr mobility.

In addition to the octet resonances there are "crystal field" scattering resonances with q = $2\pi/a$ that are strong for $\mathbf{q}$ along (0,1) or (-1, 0). In the filamentary picture these would correspond to an "L" ($\pi/2$) turn or to a dead end ($\pi$) turn, respectively. These filamentary turns could also account for some of the many narrow vibronic bands observed by ellipsometric synchrotron spectroscopy (Bernhard *et al.* 2002; Phillips 2003). Note that these peaks are as narrow and as well-defined as the octet resonances, yet they are unrelated to the Fermi surface. Their narrow peak width again is determined by the large nanodomain diameter. This suggests that filamentary geometrical effects may be at least as, and quite possibly much more, important a factor in generating the octet resonances as density of states effects.

It is noticeable (Fig. 4a (McElroy *et al.* 2002)) that the $\varepsilon(\mathbf{k})$ curves inferred from the tunneling octet analysis are offset towards higher hole energies from the center of the band of Fermi surface lines inferred from ARPES experiments. This offset is the result of broadening of Bloch-like states into wave packets by filamentary percolation through the nanodomain network. The scale is the same as that set by the $\xi \sim 1.5$ nm coherence length of the nanodomain broadening of the q resonances.

## 4. The D-Wave Surface Gap

The d-wave surface gap has been the basis of many microscopic calculations, but it has been discussed only as a *description* of experimental data. One can understand the *origin* of the large-scale surface d-wave anisotropic gap if one realizes that all observed anisotropies arise from the octet states in the weak-link filamentary segments in the $CuO_2$ planes, while all strong electron-phonon interactions occur in filamentary segments containing dopants that lie outside these planes. Let the set of electronic states near $E_F$ in the $CuO_2$ planes be represented by $\mathfrak{R}$, while the corresponding set that lies outside these planes is represented by $\aleph$. The complete set of electronic states near $E_F$ is represented by $\mathfrak{R} \oplus \aleph$. Filamentary states lie in the space $\mathfrak{R} \oplus \aleph$, but because $\mathfrak{R}$ is only weakly disordered, while $\aleph$ is strongly disordered, experimental data tend to resolve structure only in $\mathfrak{R}$. As a result, there have been many microscopic models that assume that only the easily observed states in $\mathfrak{R}$ are important to HTSC, and that these can be described successfully by the EMA and Fermi liquid models that are based only on the structure of the $CuO_2$ plane.

The LO phonon anomalies, measured by neutron scattering, reviewed in (Phillips 2001, Phillips and Jung 2002b), suggest that the dopants lie primarily *outside* the $CuO_2$ planes and are strongly correlated as second nearest neighbors in [10] and [01] directions, over an angular range of perhaps $\delta\Theta \sim$ 10-15°, consistent with primarily second neighbor correlations. The superconductive gap $2\Delta$ reflected in the octet dispersion relation (1) is not the average gap of the filament in $\Re \oplus \aleph$ space. Instead it represents the average [10] bipolaron (Mihailovic *et al.* 2002) nonplanar filamentary gap amplitude $\Delta_f$ *projected* onto the octet state Cooper pair amplitude $\Delta_p \sim \psi\psi$ in $\Re$ space, the weak-link $CuO_2$ planes. [There is a helpful analogy here for readers who are familiar with the Penrose projection scheme (Elser and Henley 1985) for constructing quasicrystals. The difference is that the Penrose projection concerns atomic probabilities, while here wave function amplitudes are projected.]

The strongest filamentary electron-phonon interactions can be regarded as occurring in a confined one-dimensional subspace defined by LO polarization vectors **u** tangent to the zigzag filamentary path. Electron-phonon interactions in such confined geometries have been discussed extensively in semiconductors. The key question is whether the confining potential is static or moves with the phonon. If the confining potential moves with the phonon, there is no inelastic electron-phonon scattering, as one can show by making a unitary transformation to the frame of reference that moves with the phonon (Schmid 1973; Sergeev and Mitin 2001), altering phonon drag, an effect which is large in one dimensional subspaces such as fullerene ropes (Romero *et al.* 2002). If this assumption is correct for HTSC, the LO electron-phonon interaction is used to construct a polaron, and for the Cooper pair the corresponding bipolaron share LO phonons. This strong interaction limit implies that there will be both filamentary electrons and filamentary vibronic phonons. Strong evidence for the existence of the latter has recently been obtained by ellipsometric infrared spectroscopy (Bernhard *et al.* 2002; Phillips 2003).

This unitary transformation or *projection* depends on the amplitudes of both the electron $\psi$ and the hole $\psi$ in the Cooper pair, each of which supplies an angular factor of $\exp i\theta$. In the long wave length (Landau-Ginsberg) limit the confining effect can be represented by an effective potential of the form $A(-i\nabla\psi - \mathbf{q})^*(-i\nabla\psi - \mathbf{q})$, with $A \sim 2a^2 T_g$ where a is the planar lattice constant and $T_g$ is the formation or annealing temperature at which the glassy filamentary network is formed. (Here the LO phonon wave vector **q** is also parallel to **u**.) To form a bipolaron with the dopant-based LO phonons one should retain only the coherent or elastic part of the ψψ projection, which is Re (expi2θ), not the dissipative or inelastic part Im (expi2θ), giving a function $|\Delta| = |\cos 2\theta| = |$ ( to be used in (1).

The magical appearance of the factor of 2 here in the angular exponent is reminiscent of e* = 2e in flux quantization by Cooper pairs; moreover, both occur in the long wave length (Landau-Ginsberg) limit. However, it appears that the origins of these two factors of 2 are different. For flux quantization the factor of 2 in e* = 2e arises because of gauge invariance of Cooper pairs in conventional space. The filamentary projection to the [10] and [01] dopant subspaces is different, and is more analogous to a Penrose construction, that is, projection of Cooper pairs from three-deimensional filamentary space to $CuO_2$ planes. Of course, in the long wave length limit the two mechanisms are completely compatible, as they occur in different spaces. There is no possibility of flux or particle fractionalization when the filamentary density is low, as in underdoped cuprates, in agreement with experiment, but in disagreement with many proposals based on the EMA (Wynn *et al.* 2001).

At present one cannot say what determines the scale of Δ, although it is clear that the scale of 40 meV that enters (1) does not reflect the strength of interactions in the weak-link $CuO_2$ planes, where $\Delta_p$ is small (0 in a relaxed surface $CuO_2$ plane (Misra *et al.* 2002)); instead, it appears to be much closer to the average or even maximum value of Δ for the entire filament. There is no constant or s-wave polaron contribution to the projected coherent percolation of Δ, because the LO filamentary phonons themselves percolate only along alternating [10] and [01] directions. However, there are still many subgap states that do not contribute to the octet, and appear as the noisy background in the **q** scattering correlations. Many of these these nonpercolative states may be weakly superconductive and contribute to the superconductive specific heat anomaly, but not to the tunneling energy gap. (To avoid charging energies (Coulomb blockade) the current flows preferentially into the more extended percolative states.) This might explain why the critical magnetic field measured by specific heat (no charging effects) is significantly smaller than that obtained by tunneling (Blanchard *et al.* 2002). These data indicate that reductionist continuum models based on Landau order parameters cannot describe HTSC, and that the nanoscale spatial inhomogeneities and filamentary percolative effects described here are important even at the large length scales involved in thermodynamic quantities such as critical fields. What is clear is that the d-wave picture is an incomplete description of effects even involving only those planar states that percolate over long distances. Note that this semiclassical bipolaron *projection* is iterated many times in constructing the Wannier function for a given nanodomain (at least 7 times), which has the effect of reducing the angular broadening of $\delta\Theta \sim 10\text{-}15°$ for the LO phonons to $\delta\theta \sim 5°$ for the **q** resonances.

Readers who are accustomed to thinking in effective medium terms may feel that this discussion of projections is really unnecessary. If the d-wave gap is observed in surface experiments – ARPES or Fourier transform STM – then isn't it a physical observable, and as such can be manipulated using effective medium methods? So it would seem, but here appearances are very deceptive. The ARPES experiments themselves show that the d-wave gap is not a well-behaved (analytic) effective medium variable. They show (Bogdanov *et al.* 2000), for example, much smaller and more weakly energy-dependent peak broadenings in the gap nodal (11) direction than in the gap antinodal (10) direction, which is the exact opposite of what any theory based on the EMA would predict. This behavior is easily understood in the filamentary model (Phillips 2003). More generally, the volume of configuration space accessed by the filaments is exponentially smaller than that of effective medium configuration space, and so it would be in general extremely accidental that projections to the former smaller from the much larger latter space are analytically well behaved. In the case of the d - wave gap measured by ARPES and by Fourier transform STM, the two agree only because the latter searched a very large area with dimensions comparable to the filamentary coherence length, which also determines oscillator strengths for photonic excitations from near-gap initial electronic states. This small **q** region is actually only an exponentially small part of the entire **q** phase space explored in the dynamical electron-phonon interactions that give rise to HTSC. Such small parts of phase space do not show up in some experiments that are sensitive to the bulk density of states for a wide range of **q**, for instance, pulsed photoinduced reflectance and absorption (Mueller 2002, Mihailovic *et al.* 2002). Of course, it is the wide range of **q** that determines fundamental interaction strengths such as the Cooper pair formation energy.

While the ARPES and large-scale deconvoluted STM data indicate a planar d-wave gap, at smaller length scales, below the nanodomain length, one can measure other gaps, and these may show electron-phonon fine structure if the probe configuration is accidentally favorable (Shimada *et al*. 1995; Ohyagi *et al.* 1995), that is, if it projects the ℵ local gap outside the ℜ $CuO_2$ planes. No electron-phonon fine structure can be observed in projections from the $CuO_2$ planes even in the red squares, because the e-p interaction is weak there and is masked by pseudogap tails from adjacent blue squares.

## 5. Conclusions

The general principle that guides the construction of theoretical models of complex systems is economy of means, that is, one seeks the simplest possible model that is compatible with experiment, and that principle has determined the model discussed in this paper. (McElroy *et al.* 2002) ask, what is causing the scattering that determines the anomalously long-range modulations of the local densities of states that they observe as sub-gap resonances? It is certainly not stripe structures, or any other kind of structure resolvable by conventional diffraction from the host lattice. (Even the most sophisticated X-ray methods (Haskel *et al.* 1996) identify filamentary effects only in the context of local orthorhombicity.) In the filamentary model the answer to this question is that the resonances represent states that are bound to self-organized zigzag dopant paths, in other words, the scattering is due to self-organized dopants. (The word "impurity" should be reserved for elements such as Zn or Ni, etc., that disrupt filamentary paths. Note that the host/dopant dichotomy explains the differences between structural and electronic phase diagrams (Phillips 2002c); these differences have not been discussed in models based on the EMA.) Analysis of this model required assumptions about the narrowing effects on probability distributions of quantum percolation over large distances; these can probably be refined, but one should keep in mind the basic limitations imposed by the resemblance of the percolative network to a quantum computer (Phillips and Jung 2002b; Phillips 2002b). Progress in this direction will probably require large-scale numerical simulations. These could be carried out in stages of increasing disorder starting from Fig. 1(b), using semiclassical billiard methods, with kinetic energies not of free billiard wave packets, but taken from (1). Strongly disordered repulsive hard walls would represent the checkerboard boundaries; these have given a good account of S-N contacts (Kormányos et al. 2002) and orbital negative magnetoresistive effects (Polyakov et al. 2001) in depletion layer quantum antidot studies. (One could also use discrete BCS methods (Richardson 1963, 1977)). If the resulting octets are insufficiently robust, then self-organized dopant attractors could be added to the model. (Dopant attractors would be particularly effective at nanodomain corner joints, where they could stabilize and enhance L turns.) These might be adequate to describe qualitatively the effects of phase coherence over long distances, where one can hope that the correspondence principle would apply. Finally, should all else fail, one would be faced with constructing generalized Wannier basis states for each red nanodomain, and connecting these through joints to adjacent nanodomains by amplitude and phase matching, a very large task.

There are several amusing analogies between very complex octet-based self-organized filaments and significantly less complex, but manifestly self-organized, proteins. These are of interest because the mathematical basis for the stability of protein structures in the presence of a high level of functional activity at low energies is now well understood in the context of network stiffness percolation (Rader et al. 2002). Moreover, network glasses exhibit phase diagrams closely resembling those of HTSC (Phillips 2002b), and these diagrams are also well understood in the context of network stiffness percolation. There are two robust (rigid) structural motifs in proteins based on their peptide backbones . The first is the α helix, stabilized by hydrogen bonds between i and i + 4 peptide units, while the second is β strands (Fersht 1999). The latter are α helices that have reversed direction by folding back on themselves at what is called a β hairpin. The α helix is analogous to the momentum quartet $(\pm k_x, \pm k_y)$, the β strand is analogous to the reversed quartets $[(\pm k_x, \pm k_y)$ and $(\pm k_y, \pm k_x)]$, and the β hairpin is analogous to the corner joint. The β hairpin is also stabilized by an extra hydrogen bond. For HTSC the dopants play the same role that the hydrogen bonds do for proteins, that is, they stabilize the filamentary topological structure and make it robust against the exponentially more numerous configurational fluctuations that are thermally active at annealing temperatures. The stabilization of the α helix by hydrogen bonds between i and i + 4 peptide units is analogous to the second neighbor [10] and [01] dopant correlations that are reflected in the LO phonon dispersion anomalies. The projection of the latter from $\aleph$ space to $\Re$ space generates the planar d wave anisotropy.

Another interesting biophysical analogy involves charge transport in DNA, which may be of functional value in repairing damage. DNA is one-dimensional, and if it is randomly disordered, it will be a semiconductor. Attempts to analyze disorder in DNA using crude power-law analysis of amino sequences show weak correlations too close to random to produce substantial conductivity (Holste *et al.* 2001; Carpena *et al.* 2002), but power-law analysis exponentially underestimates the effectiveness of phase-coherent self-organization. In HTSC dopant self-organization is driven by the energy gained from screening fluctuating internal electric fields by highly conductive filaments. The remarkably robust subgap octet resonances observed by (McElroy *et al.* 2002) resemble impurity bands associated with resonant tunneling, which could explain observed charge transport in DNA (Hjort and Stafstrom 2001).

The filamentary model provides an explanation for the origin of the narrow and anomalously intense **q** resonances and the **k**-space octet states. It also *predicted* the **r**-space checkerboard. It shows how infinitely many Fermi surface **k** octets can retain their distinct individual identities in the presence of a very high irregular checkerboard **r** - space noise level, and yet share functionally similar **q** octets. It explains the *origin* of (rather than merely describes) the planar d-wave gap measured by ARPES and inferred from the octet STM structure. It shows that the origin of the d-wave gap is primarily topological, in the sense that long-range current-carrying states are bound (negative energy states, below the pseudogap) to self-organized dopant filaments; these states are not describable in terms of scattering at positive energies from any kind of individual impurities (Byers *et al.* 1993). It relates the d-wave gap anisotropy to the anisotropic phonon anomalies measured by neutron scattering, and shows that the two anisotropies are consistent. Thus it provides a *complete platform* for understanding HTSC within a semiclassical percolation model using Cooper pairs created by strong dopant-mediated electron-phonon interactions outside the $CuO_2$ planes. This platform can be used for further analysis of the behavior of quantum computers, as the octet construction has revealed a large-scale semiclassical structure that is much more robust in the presence of a high level of noise than had been previously anticipated.

I am grateful to J. C. Davis for a preprint of (McElroy *et al.* 2002) and for his patience during several clarifying discussions.

## FIGURE CAPTION

Fig. 1. (a) A regular red (superconductive gap) and blue (pseudogap) checkerboard. (b) A first distortion of (a), showing how tetragonal symmetry can be deformed into alternating orthorhombic symmetry. In these drawings the detailed structure near the corners is not shown. It is likely that the percolative dynamics near such corner joints is modified by the presence of dopant attractors.

**Fig. 1**

| B | R | B |
|---|---|---|
| R | B | R |
| B | R | B |

(a)

| B | R | B |
|---|---|---|
| R | B | R |
| B | R | B |

(b)